\documentclass[reprint, superscriptaddress, secnumarabic, amssymb, nobibnotes, aps, pra]{revtex4-1}

\newcommand{\musr}{$\mu$SR} 
\usepackage{graphicx}
\usepackage{epstopdf}
\usepackage[T1]{fontenc}
\usepackage[latin9]{inputenc}
\usepackage{amsbsy}
\usepackage{gensymb}
\usepackage[T1]{fontenc}
\usepackage[latin9]{inputenc}
\usepackage{amsmath}
\usepackage{amssymb}
\usepackage{bbm}
\usepackage{braket}
\usepackage{xcolor}
\usepackage{mathrsfs}
\allowdisplaybreaks
\usepackage{graphicx}
\usepackage[colorlinks=true]{hyperref}  
\hypersetup{
    bookmarks=true,         % show bookmarks bar?
    unicode=false,          % non-Latin characters 
    pdftoolbar=true,        % show Acrobat
    pdfmenubar=true,        % show Acrobat 
    pdffitwindow=false,     % window fit to page when opened
    pdfstartview={FitH},    % fits the width of the page to the window
    pdftitle={Time-reversal-symmetry breaking and unconventional pairing in the noncentrosymmetric superconductor La$_{7}$Rh$_{3}$ probed by \musr},    % title
    pdfauthor={D. Singh, M. Scheurer, A. D. Hillier, R. P. Singh},     % author
    pdfsubject={},   % subject of the document
    pdfcreator={},   % creator of the document
    pdfproducer={}, % producer of the document
    pdfkeywords={} {} {}, % list of keywords
    pdfnewwindow=true,      % links in new window
    colorlinks=true,       % false: boxed links; true: colored links
    linkcolor=blue, %red,          % color of internal links (change box color with linkbordercolor)
    citecolor=blue,        % color of links to bibliography
    filecolor=magenta,      % color of file links
    urlcolor=blue           % color of external links
} 
\usepackage[normalem]{ulem}

\newcommand{\equref}[1]{Eq.~(\ref{#1})}
\newcommand{\equsref}[2]{Eqs.~(\ref{#1}) and (\ref{#2})}

\newcommand{\figref}[1]{Fig.~\ref{#1}}
\newcommand{\refcite}[1]{Ref.~\onlinecite{#1}}

\newcommand{\tableref}[1]{Table~\ref{#1}}
\newcommand{\appref}[1]{Appendix~\ref{#1}}
\newcommand{\pdagger}{{\phantom{\dagger}}}

\renewcommand{\approx}{\simeq}

\renewcommand{\vec}[1]{\boldsymbol{#1}}

\begin{document}
\title{Time-reversal-symmetry breaking and unconventional pairing in the noncentrosymmetric superconductor La$_{7}$Rh$_{3}$}
\author{D.~Singh}
\affiliation{Department of Physics, Indian Institute of Science Education and Research Bhopal, Bhopal, 462066, India}
\affiliation{ISIS Facility, STFC Rutherford Appleton Laboratory, Harwell Science and Innovation Campus, Oxfordshire, OX11 0QX, UK}
\author{M.~S.~Scheurer}
\affiliation{Department of Physics, Harvard University, Cambridge MA 02138, USA}
\affiliation{Institute for Theoretical Physics, University of Innsbruck, A-6020 Innsbruck, Austria}
\author{A.~D.~Hillier}
\affiliation{ISIS Facility, STFC Rutherford Appleton Laboratory, Harwell Science and Innovation Campus, Oxfordshire, OX11 0QX, UK}
\author{D.~T.~Adroja}
\affiliation{ISIS Facility, STFC Rutherford Appleton Laboratory, Harwell Science and Innovation Campus, Oxfordshire, OX11 0QX, UK}
\author{R.~P.~Singh}
\email[]{rpsingh@iiserb.ac.in}
\affiliation{Department of Physics, Indian Institute of Science Education and Research Bhopal, Bhopal, 462066, India}
\date{\today}
\begin{abstract}
\begin{flushleft}
\end{flushleft}
Noncentrosymmetric superconductors have sparked significant research interests due to their exciting properties, such as the admixture of spin-singlet and spin-triplet pairing. 
Here we report on the  \musr~and thermodynamic measurements on the noncentrosymmetric superconductor La$_{7}$Rh$_{3}$ which show an isotropic superconducting gap but also spontaneous time-reversal-symmetry breaking occurring at the onset of superconductivity.
We show that our results pose severe constraints on any microscopic theory of superconductivity in this system. 
A symmetry analysis identifies ground states compatible with time-reversal-symmetry breaking, and the resulting gap functions are discussed. Furthermore, general energetic considerations indicate the relevance of electron-electron interactions for the pairing mechanism, in accordance with hints of spin-fluctuations revealed in susceptibility measurements.  
\end{abstract}

\maketitle

\section{INTRODUCTION}

The symmetry of the order parameter plays a pivotal role in determining the nature of the superconducting ground state \cite{SigristReview}. The key symmetries that are associated with superconductivity are spatial inversion, gauge and time-reversal symmetry (TRS). TRS is intimately related to superconductivity as Cooper pairs are built from Kramers partners \cite{PW1}. The superconducting ground state in systems which exhibit inversion symmetry in the crystal structure (centrosymmetric superconductors) can be expressed distinctly via the parity of the Cooper pair state. If the spin part of the Cooper pairs is singlet, then the orbital part corresponds to even parity whereas the spin-triplet pair state requires odd-parity orbital wave functions. However, a remarkably different situation occurs in a noncentrosymmetric superconductor (NCS) with spin-orbit coupling:
as parity is not a good quantum number in the normal state, the absence of inversion symmetry leads to the mixing of singlet and triplet pairing \cite{rashba,Bauer2004,PA,EBA}. Together with the antisymmetric spin-orbit coupling, which removes the spin degeneracy of the electronic bands \cite{EI}, this leads to the emergence of many exciting superconducting properties \cite{rashba,Bauer2004,PA,EBA,JC,AB,GB,TP,Caviglia,Qi,Oxide,Coleman,PairBr}.

A particularly interesting and rare phenomenon is TRS-breaking superconductivity. Exclusivity of TRS breaking can be adjudged by the fact that to date only a few superconductors were found to break TRS, e.g., Sr$_{2}$RuO$_{4}$ \cite{GML,JXY}, UPt$_{3}$  and (U,Th)Be$_{13}$ \cite{GM,PD,WH,RHH}, (Pr,La)(Os,Ru)$_{4}$Sb$_{12}$ \cite{YA,LS}, PrPt$_{4}$Ge$_{12}$ \cite{AM}, LaNiGa$_{2}$ \cite{ADH}, Lu$_{5}$Rh$_{6}$Sn$_{18}$ \cite{LOS}, Ba$_{0.27}$K$_{0.73}$Fe$_{2}$As$_{2}$ \cite{VGP}. In systems with broken TRS, the non-zero moments of the Cooper pairs locally align to induce an extremely small spontaneous internal field (0.01$\,\mu_B$), which is difficult to be detected by most measurement techniques. Muon spin rotation and relaxation ($\mu$SR) \cite{AS,AY} is a technique which is extremely sensitive to such tiny changes in internal fields and  
can measure the effect of TRS breaking in these exotic superconductors. Moreover, the symmetry of the superconducting gap function can also be inferred from $\mu$SR by measuring the temperature dependence of the magnetic penetration depth.

The search for NCSs with broken TRS is driven by the fundamental interest in the interplay of inversion symmetry of the crystal structure, spin-orbit coupling, and TRS. However, it has only been  observed in a few compounds, e.g., LaNiC$_{2}$ \cite{ADJ}, Re$_{6}$Zr \cite{RPS}, Re$_{6}$Hf \cite{DSJ}, SrPtAs \cite{PKB}, La$_{7}$Ir$_{3}$ \cite{JAT} even though many NCSs have been studied so far \cite{RAD,VK,MS,VKD,TKF,PAM,EC}.

In this work, we report the $\mu$SR study of the NCS La$_{7}$Rh$_{3}$. Zero-field  $\mu$SR reveals that spontaneous magnetic fields develop at the superconducting transition temperature, identifying the presence of spontaneous TRS breaking. Furthermore, the temperature dependence of the magnetic penetration depth determined from the transverse-field muon measurements indicates a nodeless, isotropic gap. We show that our findings impose strong restrictions on the possible microscopic superconducting order parameters and indicate that the mechanism of superconductivity must be unconventional, i.e., driven by electron-electron interactions.

\section{Experimental details}
Single phase polycrystalline samples of La$_{7}$Rh$_{3}$ were prepared by melting together a stoichiometric mixture of La (99.95$\%$, Alfa Aesar) and Rh (99.99$\%$, Alfa Aesar) in an arc furnace under a high-purity argon gas atmosphere on a water-cooled copper hearth. The sample buttons were melted and flipped several times to ensure phase homogeneity with negligible weight loss. To verify the phase purity we  performed room temperature (RT) powder x-ray diffraction (XRD) using a X'pert PANalytical diffractometer (Cu-K$_{\alpha 1}$ radiation, $\lambda$ = 1.540598 \text{\AA}). A Quantum Design superconducting quantum interference device (SQUID) and physical property measurement system (PPMS) were used to measure the magnetization and specific heat of La$_{7}$Rh$_{3}$.

In order to probe the superconducting ground state locally and to further understand the superconducting gap structure of our sample, $\mu$SR experiments were carried out at the ISIS Neutron and Muon facility, in STFC Rutherford Appleton Laboratory, United Kingdom,
using the MUSR spectrometer. A full description of the $\mu$SR technique may be found in Ref. \cite{MSM}. $\mu$SR measurements in transverse-field (TF), zero-field (ZF), and longitudinal-field (LF) configurations are used to probe the flux line lattice (FLL) and TRS breaking. The powdered sample of La$_{7}$Rh$_{3}$ was mounted on a high purity silver sample holder which is then placed in a dilution fridge, which can operate in the temperature range from $40\,\textrm{mK}$ to $4\,\textrm{K}$.

\begin{figure}
\includegraphics[width=1.0\columnwidth]{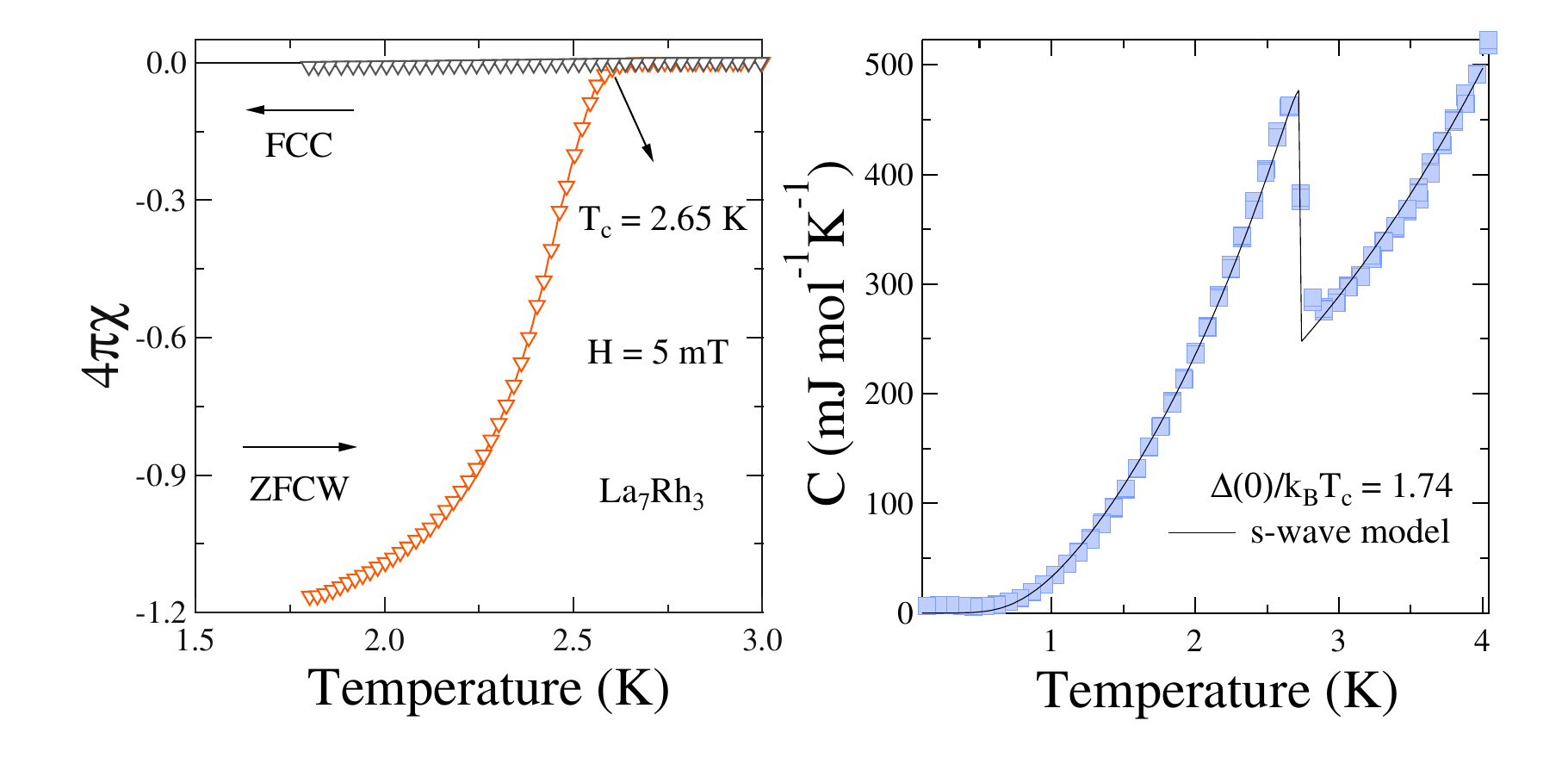}
\caption{\label{Fig1:xrd} (color online) (a) Superconductivity appears around $T_{c}$ = 2.65 K with the onset of strong diamagnetic signal in the zero-field cooled warming (ZFCW) and field cooled cooling (FCC) magnetization measurements. (b) Low temperature specific heat data in the superconducting regime fits well for BCS s-wave model for a fitting parameter $\Delta (0)/k_{B}T_{c}$ = 1.74.}
\end{figure}

\section{Results}
La$_{7}$Rh$_{3}$ crystallizes in a hexagonal structure with space group $P6_{3}mc$ (No.~186) \cite{TH}, which we confirmed for our sample by performing room-temperature powder x-ray diffraction (XRD) measurements.
The lattice constants are $a = 10.203 \pm 0.002$ \text{\AA} and $c = 6.505 \pm 0.002$ \text{\AA}. Very importantly, no impurity phases were observed.
The superconducting transition temperature was found to be $T_{c}$ = 2.65 $\pm$ 0.02 K [\figref{Fig1:xrd}(a)] from magnetization measurements, which is in good agreement with the published literature \cite{TH}. The lower and upper critical fields $H_{c1}(0)$ and $H_{c2}(0)$ were obtained using Ginzburg-Landau expressions, which yields $H_{c1}(0)$ = 2.51 $\pm$ 0.02 mT and $H_{c2}(0)$ = 1.02 $\pm$ 0.03 T. The upper critical field is thus significantly smaller than the Pauli paramagnetic limiting field $H_{c2}^{p}$(0) = 1.86$T_{c}$ = 4.93~T. This indicates a significant singlet component rather than dominant spin triplet. Using the standard relations given in Ref. \cite{MTI}, we obtain a coherence length of $\xi_{GL}$(0) = 179 \text{\AA} and a penetration depth $\lambda_{GL}$(0) = 4620 \text{\AA}. The normalized specific heat jump at $T_{c}$ is $\Delta C_{el}/\gamma_{n}T_{c}$ = 1.38 $\pm$ 0.02, which is close to the value reported earlier \cite{PP} and to the BCS value 1.43, indicating weakly-coupled superconductivity in La$_{7}$Rh$_{3}$. The specific heat data in the superconducting state below $T_{c}$ fits well to that of a superconductor with a single, isotropic gap [\figref{Fig1:xrd}(b)], for $\Delta (0)/k_{B}T_{c}$ = 1.74.

\begin{figure}
\includegraphics[width=1.0\columnwidth]{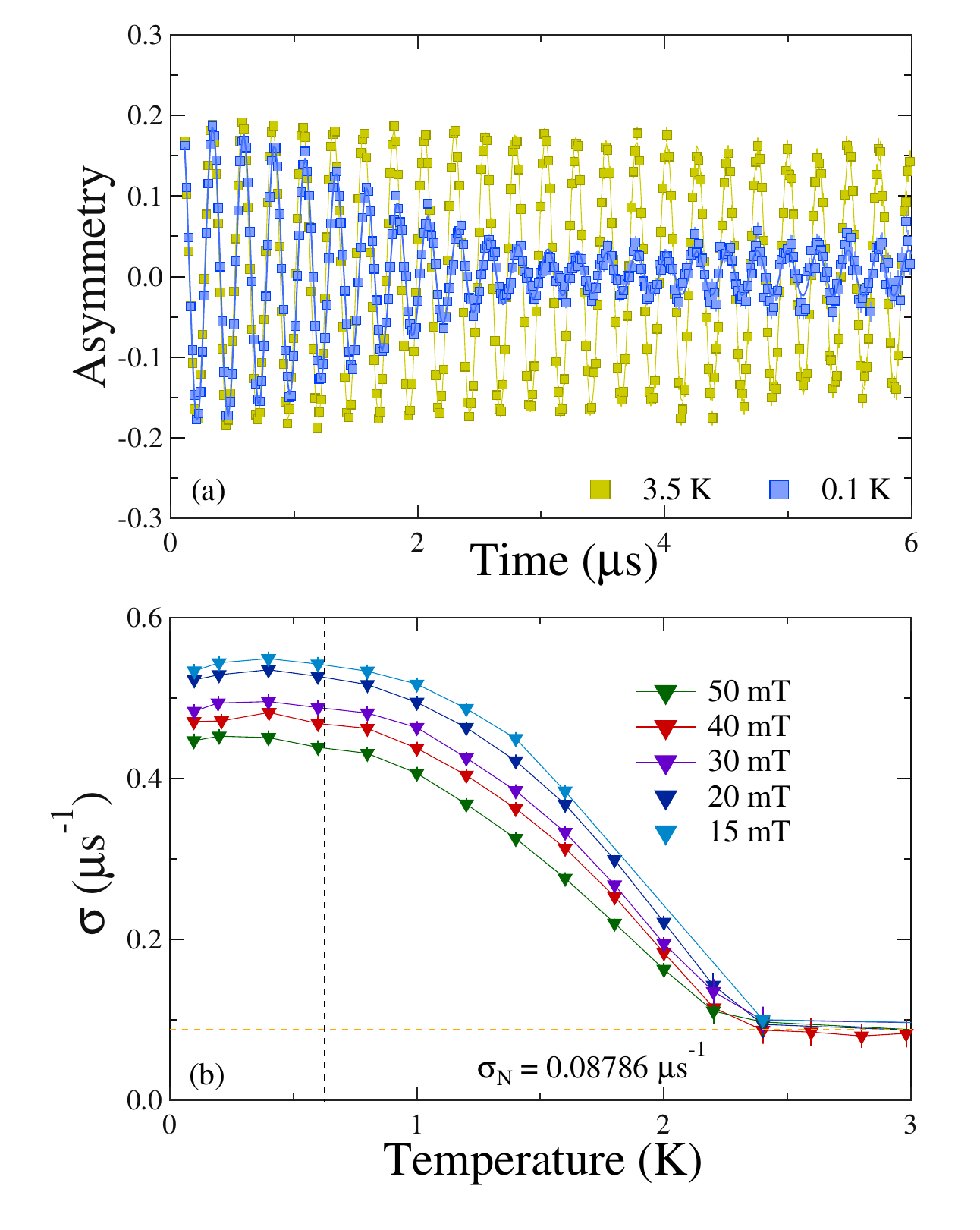}
\caption{\label{Fig2:hc2} (color online)(a) TF-$\mu$SR spectra collected in an applied magnetic field of 30 mT at temperature 3.5 K ($>T_{c}$)  and 0.1 K ($<T_{c}$). The solid lines are fits using \equref{eqn1:Tranf}. (b) Temperature dependence of the muon-spin relaxation rate at different applied magnetic fields from 15 mT to 50 mT.}
\end{figure}

 Transverse-field $\mu$SR (TF-$\mu$SR) measurements provide detailed information on the nature of the superconducting gap. The TF-$\mu$SR measurements were performed in the superconducting mixed state in applied fields between 15 mT $\le H \le$ 50 mT, well above the $H_{c1}(0)$ of this material. The data were collected in the field-cooled mode, where a field of $H = 30$~mT was applied perpendicular to the initial muon spin direction from a temperature above the transition temperature to the base temperature, in order to establish a well ordered FLL in the mixed state. Figure \ref{Fig2:hc2}(a) shows the signal in the normal state ($T = 3.5 \,\mathrm{K} >  T_{c}$) where the depolarization rate is small, attributed to the homogeneous field distribution throughout the sample. The significant depolarization rate in the superconducting state ($T = 0.1 \,\mathrm{K} < T_{c}$) is due to an inhomogeneous field distribution of the FLL.
 
The TF-$\mu$SR asymmetry spectra can be described by the sum of cosines, each damped with a Gaussian relaxation term \cite{AMR,MWA}:
\begin{equation}
G_\mathrm{TF}(t) = \sum_{i=1}^N A_{i}\exp\left(-\frac{1}{2}\sigma_i^2t^2\right)\cos(\gamma_\mu B_it+\phi),
\label{eqn1:Tranf}
\end{equation}
where $A_{i}$ is the initial asymmetry, $\sigma_i$ the Gaussian relaxation rate, $\gamma_{\mu}/2\pi$ = 135.5 MHz/T the muon gyromagnetic ratio, $\phi$ the common phase offset, and $B_i$ is the first moment for the $i$th component of the field distribution. We found that the asymmetry spectra of our sample can best be described by two oscillating functions ($N=2$), where the second component of the depolarization rate was fixed to zero ($\sigma_{2} = 0$) which accounts for the non-depolarizing muons that stop in the silver sample holder. Additional Gaussian terms were also tried, but no improvement in the fit quality was obtained. The field distribution in the mixed state of a superconductor is broadened by the presence of static fields arising from the nuclear moments.

\begin{figure}
\includegraphics[width=1.0\columnwidth]{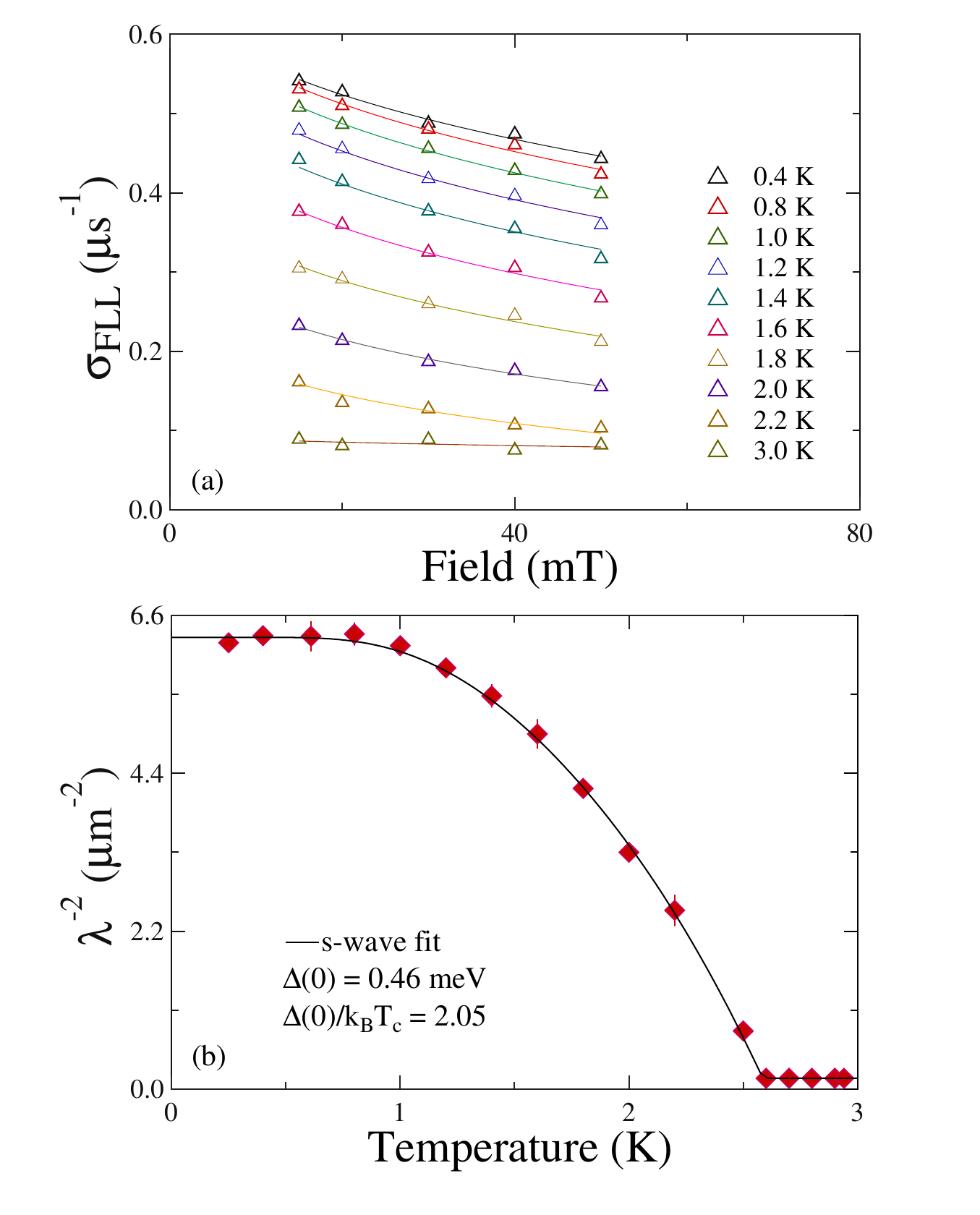}
\caption{\label{Fig3:TF} (color online)  (a) Muon spin depolarization rate as a function of field at various temperatures. The data were fitted using \equref{eqn3:sigmaH} to extract the temperature dependence of the inverse magnetic penetration depth squared. (b) Temperature dependence of $\lambda^{-2}$ is shown where the solid line represents the best fit using \equref{eqn4:LPD}.}
\end{figure}
	
The temperature dependence of the muon-spin relaxation rate $\sigma = \sigma_{1} $ has been determined for different applied magnetic fields and is displayed in \figref{Fig2:hc2}(b). The background nuclear dipolar relaxation rate $\sigma_{\mathrm{N}}$ obtained from the spectra above $T_{c}$ was assumed to be temperature independent over the temperature range of study. It is shown by the dotted horizontal orange line in \figref{Fig2:hc2}(b), with $\sigma_{\mathrm{N}}$ = 0.08786 $\pm$ 0.0032 $\mu$s$^{-1}$, which is then subtracted quadratically from the total sample relaxation rate $\sigma$ to extract the superconducting contribution to the muon-spin relaxation rate $\sigma_{\mathrm{FLL}}=\sqrt{\sigma^{2} - \sigma_{\mathrm{N}}^{2}}$.
Figure \ref{Fig3:TF}(a) shows the field dependence of the depolarization rate $\sigma_{\text{FLL}}(H)$ evaluated using isothermal cuts of the $\sigma(T)$ data sets in \figref{Fig2:hc2}(b). In an isotropic type-II superconductor with a hexagonal Abrikosov vortex lattice the magnetic  penetration depth $\lambda$ is related to $\sigma_{\mathrm{FLL}}$ by the equation \cite{EH}:
\begin{equation}
\sigma_{\mathrm{FLL}}(\mu s^{-1}) = 4.854 \times 10^{4}(1-h)[1+1.21(1-\sqrt{h})^{3}]\lambda^{-2} , 
\label{eqn3:sigmaH}
\end{equation}
where $h = H/H_{c2}$ is the reduced field. The temperature dependence of $\lambda^{-2}$ is displayed in \figref{Fig3:TF}(b), which was extracted by fitting \equref{eqn3:sigmaH} to the data presented in \figref{Fig3:TF}(b) taking into account the temperature dependence of the upper critical field $H_{c2}$. Note that  $\lambda^{-2}(T)$ is nearly constant below $T_{c}$/3 $\approx$ 0.88 K. This suggests the absence of low-lying excitations and is indicative of a nodeless superconducting gap at the Fermi surface. This is verified by the temperature dependence of the London magnetic penetration depth $\lambda(T)$ within the local London approximation for an s-wave BCS superconductor in the clean limit using the following expression:
\begin{equation}
\frac{\lambda^{-2}(T)}{\lambda^{-2}(0)} = 1+2\int_{\Delta (T)}^{\infty}\left(\frac{\partial f}{\partial E}\right) \frac{E dE}{\sqrt{E^{2}-\Delta^{2}(T)}},
\label{eqn4:LPD}
\end{equation}
where $f({E}) = [\exp({E}/k_{B}T)+1]^{-1}$ is the Fermi function and $\Delta({T})/\Delta(0) = \tanh[1.82(1.018({T_c/T}-1))^{0.51}]$ is an approximate solution to the BCS gap equation \cite{AproxSolGap}. The above discussed model fits ideally [see \figref{Fig3:TF}(b)] for the fitted value of the energy gap $\Delta (0)$ = 0.462 $\pm$ 0.004 meV, which yields the BCS parameter $\Delta (0)/k_{B}T_{c} = 2.02 \pm 0.02$. This is larger than the value of 1.76 expected from weak-coupling BCS theory and what we found from specific heat above; a similar discrepancy has been reported for La$_7$Ir$_3$ \cite{JAT,DFT}.
\begin{figure*}[t]
\includegraphics[width=2.0\columnwidth]{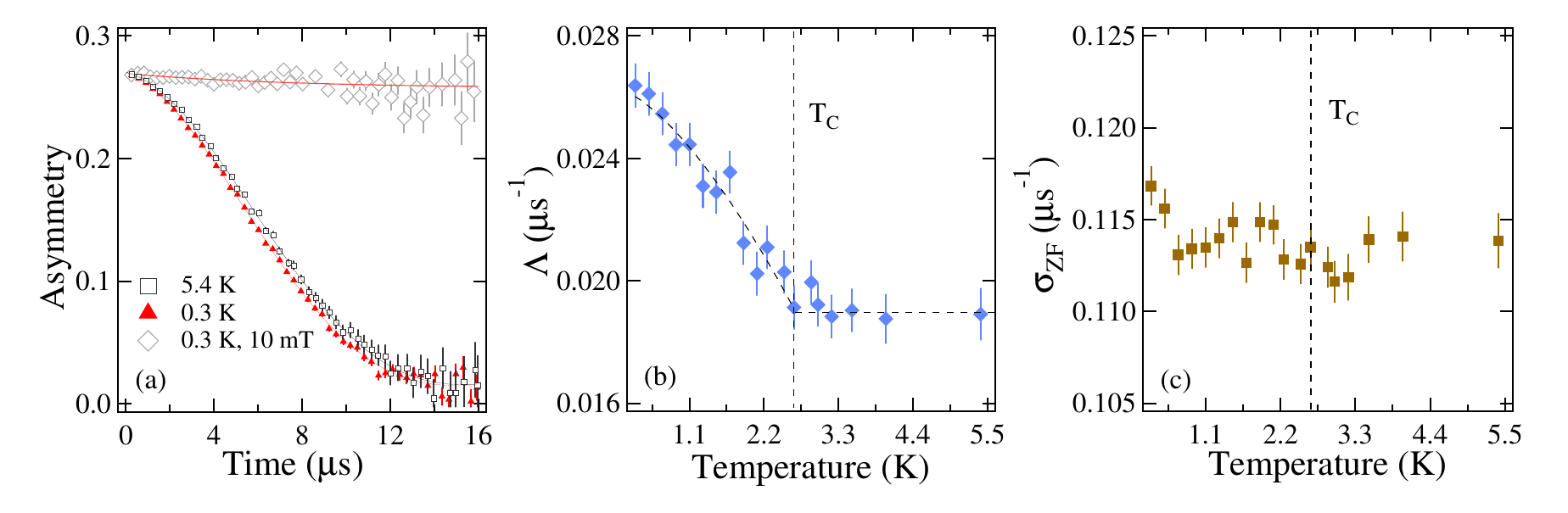}
\caption{\label{Fig4:TF} (color online) (a) ZF-$\mu$SR spectra at 0.1 K and 3.5 K. The orange circles refer to measurements in the presence of a small LF of 10 mT. (b) The temperature dependence of the electronic relaxation rate $\Lambda$ shows a systematic increase below T = 2.64 K, which is close to $T_{c}$. (c) The temperature dependence of the nuclear relaxation rate $\sigma_{\text{ZF}}$ shows no appreciable change at $T_{c}$.}
\end{figure*}

Zero-field muon spin relaxation (ZF-$\mu$SR) measurements are carried out in order to look for the tiny spontaneous magnetization associated with TRS breaking in the superconducting state. The time evolution of the relaxation spectra was collected below ($T$ = 0.3 K) and above ($T$ = 5.4 K) the transition temperature ($T_{c}$ = 2.65 K) as displayed in \figref{Fig4:TF}(a). There are no visible oscillatory components in the spectra, ruling out the presence of any ordered magnetic structure. As \figref{Fig4:TF}(a) illustrates, there is stronger relaxation below the superconducting transition temperature $T_{c}$, which suggests the presence of internal magnetic fields in the superconducting state.\\
The ZF-$\mu$SR in the absence of atomic moments and muon diffusion is best described by the Gaussian Kubo-Toyabe (KT) function \cite{RS} 
\begin{equation}
G_{\mathrm{KT}}(t) = \frac{1}{3}+\frac{2}{3}(1-\sigma^{2}_{\mathrm{ZF}}t^{2})\mathrm{exp}\left(\frac{-\sigma^{2}_{\mathrm{ZF}}t^{2}}{2}\right) ,
\label{eqn5:zf}
\end{equation} 
where $\sigma_{\mathrm{ZF}}$ denotes the relaxation due to static, randomly oriented local fields associated with the nuclear moments at the muon site.

The spectra obtained for La$_{7}$Rh$_{3}$ comprise additional relaxation signals below $T_{c}$, that can be described by the function
\begin{equation}
A(t) = A_{1}G_{\mathrm{KT}}(t)\mathrm{exp}(-\Lambda t)+A_{\mathrm{BG}} ,
\label{eqn6:tay}
\end{equation} 
where $A_{1}$ is the initial asymmetry, $A_{\mathrm{BG}}$ is the time independent background contribution from the muons stopped in the sample holder whereas the exponential term ($\mathrm{exp}(-\Lambda t)$) accounts for the presence of additional electronic relaxation channels. 
The parameters $A_{1}$ and $A_{\mathrm{BG}}$ are found to be approximately temperature independent with statistical average value $A_{1}$ =  0.26185(4),  $A_{\mathrm{BG}}$ = 0.00865(4). The nuclear depolarization rate $\sigma_{\mathrm{ZF}}$ was also found to be approximately temperature independent as displayed in \figref{Fig4:TF}(c). Interestingly, the electronic relaxation rate parameter $\Lambda$ shows a clear increase below the temperature $T = 2.64 $ K [see \figref{Fig4:TF}(b)], which is close to the superconducting transition temperature.  
Such a systematic increase in $\Lambda$ below $T_{c}$ was also identified in compounds like Sr$_{2}$RuO$_{4}$ \cite{GML}, LaNiC$_{2}$ \cite{ADJ} and the locally noncentrosymmetric compound SrPtAs \cite{PKB} by $\mu$SR measurements, where the particular behavior was attributed to the formation of spontaneous magnetic fields below $T_{c}$, which in turn indicates TRS breaking in these compounds. These observations clearly suggest that TRS is broken in the superconducting state of La$_{7}$Rh$_{3}$.

To eliminate the possibility that the above signal is due to extrinsic effects such as impurities, we also performed LF-$\mu$SR measurements. As shown by the grey markers in \figref{Fig4:TF}(a), the magnetic field of 10~mT was sufficient to fully decouple the muons from the electronic relaxation channel. This indicates that the associated magnetic fields are in fact static or quasistatic on the time scale of the muon precession. This further provides unambiguous evidence for TRS breaking in La$_{7}$Rh$_{3}$ in the superconducting state. The increase $\Delta\Lambda$ in the relaxation channel below $T_{c}$ for most of the NCS superconductors with broken TRS was found between 0.005 $\mu$s$^{-1}$and 0.05 $\mu$s$^{-1}$ \cite{GML,GM,ADJ,RPS,PKB,JAT,MNW}. In our case, $\Delta\Lambda\approx$ 0.007 $\mu s^{-1}$, suggesting a smaller TRS breaking field.

%\vspace{1.3em}

\section{Discussion}

Let us next discuss the implications of our experimental findings for the possible microscopic superconducting order parameters and the interactions driving the superconducting instability. The point group $C_{6v}$ of the normal state above $T_c$ together with the expected sizable spin-orbit coupling resulting from the large atomic numbers of La and Rh allows for 10 distinct pairing states -- four associated with the four different 1D ($d_n=1$) irreducible representations (IRs) and three with each of the two 2D ($d_n=2$) IRs of $C_{6v}$. In centrosymmetric systems, the superconducting order parameter is usually expressed in the (pseudo)spin basis, $\Delta_{\vec{k},\alpha\beta}$ with $\alpha$, $\beta$ denoting the (pseudo)spin of the two electrons forming a Cooper pair and $\vec{k}$ their relative momentum. 
This is not a good basis in systems without a center of inversion, like La$_{7}$Rh$_{3}$, where the spin-degeneracy of the electronic bands is removed at generic momenta $\vec{k}$ and it is physically more insightful to describe $\Delta_{\vec{k},\alpha\beta}$ in the band basis: at low energies, the superconducting order parameter is described by a single complex function $\widetilde{\Delta}_{\vec{k}a} \in \mathbbm{C}$ for each non-degenerate band $a$ of the normal state which is obtained by projecting $\Delta_{\vec{k},\alpha\beta}$ on the respective band \cite{MSS2}. If the order parameter transforms under the IR $n$, it holds $\widetilde{\Delta}_{\vec{k}a} = \sum_{\mu=1}^{d_n} \eta_\mu\varphi^\mu_{na}(\vec{k})$ with complex coefficients $\eta_\mu$ and continuous, real-valued basis functions $\varphi^\mu_{na}(\vec{k})$ transforming under $n$, see \appref{OrderParameterInBandBasis} for details.

As all IRs of $C_{6v}$ are real, it holds $\eta_\mu\rightarrow \left(\eta_\mu\right)^*$ under time-reversal and, hence, only the two 2D IRs, $E_1$ and $E_2$, are consistent with the observation of broken TRS. Furthermore, by performing a Ginzburg-Landau analysis for $E_1$ and $E_2$, one finds that only the three discrete configurations $(\eta_1,\eta_2)=(1,0),\,(0,1),\,(1,i)$ can arise. Only the third option is consistent with the broken TRS which reduces the number of 10 possible pairing states to the two remaining states $E_1(1,i)$ and $E_2(1,i)$.

\begin{figure}
\includegraphics[width=\columnwidth]{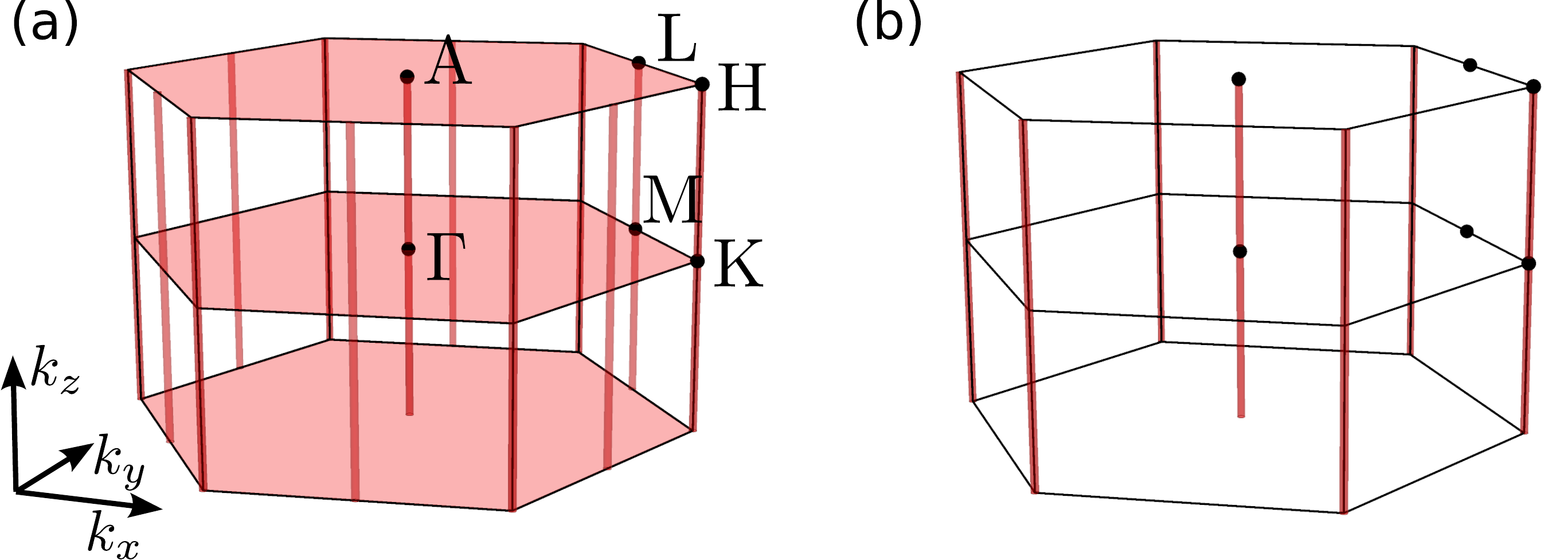}
\caption{\label{SymmProtectZeros}The location of symmetry-imposed zeros of the single-band gap function $|\widetilde{\Delta}_{\vec{k} a}|$ in the Brillouin zone is shown in red for the TRS-breaking candidate states $E_1(1,i)$ and $E_2(1,i)$ in (a) and (b), respectively. Nondegenerate Fermi surfaces crossing these lines/planes lead to symmetry-protected point/line nodes.
Bands are expected to be nondegenerate for generic momentum points (invariant only under the identity operation of the point group), including generic points in the planes $k_z=0,\pi/c$; however, at the high-symmetry lines parallel to the $k_z$ axis and going through the indicated high-symmetry points, this will not necessarily be the case.}
\end{figure}

To further constrain the order parameter, let us next take into account the implications of our observation of a fully established, nodeless, superconducting gap. 
Being odd under two-fold rotation $C^z_2$ along the $z$ axis, the basis functions of the $E_1$ state satisfy $\varphi^\mu_{E_1a}(\vec{k})=-\varphi^\mu_{E_1a}(-k_x,-k_y,k_z)=-\varphi^\mu_{E_1a}(k_x,k_y,-k_z)$, where we used the constraint $\varphi_{na}^\mu({\vec{k}})=\varphi_{na}^\mu({-\vec{k}})$ resulting from TRS and Fermi statistics \cite{MSS2}. Consequently, the gap function $|\widetilde{\Delta}_{\vec{k}a}|$ of the $E_1(1,i)$ state necessarily vanishes on the planes $k_z=0,\pi/c$ which, in turn, will lead to line nodes for any Fermi surface crossing these planes. Note that this does not contradict Blount's theorem \cite{Blount} as we consider a noncentrosymmetric system with nondegenerate Fermi surfaces.
The $C^z_2$-even $E_2(1,i)$ state does not exhibit planes of zeros of $|\widetilde{\Delta}_{\vec{k}a}|$, and is only forced to vanish at high-symmetry points in the Brillouin zone (see \figref{SymmProtectZeros} and \appref{SymmetryProtectedZeros} for more details) where the assumption of having nondegenerate bands is not satisfied any more. We thus conclude that $E_2(1,i)$ is the most natural pairing state based on our experimental findings. However, the detailed microscopic form of the $E_2(1,i)$ order parameter, why it effectively behaves like an isotropically gapped superconductor, in particular in the vicinity of the aforementioned high-symmetry points, or whether fine-tuning/approximate symmetries (leading, e.g., to two nearly degenerate superconducting transitions or accidentally degenerate bands) are necessary remain open questions. Given the multitude of atoms in the unit cell, the phenomenological real-space picture of \refcite{AnnettNewProposal} might allow for useful insights. 

Our experimental results also provide important information about the mechanism of superconductivity: taking advantage of general mathematical properties of Eliashberg equations for noncentrosymmetric superconductors, it was shown in \refcite{MSS1} that electron-phonon coupling alone cannot give rise to TRS-breaking superconductivity. This indicates that the superconductivity of La$_{7}$Rh$_{3}$ must be crucially driven by electron-electron interactions (``unconventional pairing''). 
Furthermore, the results of \refcite{MSS1} imply that the collective electronic mode $\phi$ providing the ``pairing glue'' must be time-reversal odd, i.e., its condensation $\braket{\phi}\neq 0$ breaks time-reversal symmetry. This is, e.g., the case for spin fluctuations, signs of which have been reported for La$_{7}$Rh$_{3}$ in \refcite{PP}. The relevance of electron-electron interactions in the system is corroborated by the large value $R_w \approx 2.97$ of the Wilson ratio extracted from our measurements (see \appref{WilsonRatio}). In general, susceptibility data and the calculation of the Wilson ratio give signatures of electron-electron correlation; however, it is not sufficient to draw a definite conclusion. Therefore, specific heat measurements down to low temperatures at various high applied magnetic fields coupled with the density of state calculations, are indeed required to understand the strength of the correlation in the system. 

\section{Conclusion}
To summarize, our results indicate that the superconducting ground state in La$_{7}$Rh$_{3}$ breaks TRS while exhibiting an isotropic gap, similar to that of La$_{7}$Ir$_{3}$.  
A symmetry analysis shows that there are only two distinct candidate pairing phases consistent with TRS breaking, one of which gives rise to nodal lines and, hence, is disfavored by our observations. General energetic considerations indicate an unconventional pairing mechanism.    
Further experimental work on single crystals, coupled with theoretical studies, is required to fully unravel the microscopic nature of superconductivity in this important family of materials.

\begin{acknowledgments}
R.~P.~S.~acknowledges Science and Engineering Research Board, Government of India for the Core Research Grant CRG/2019/0010282. M.~S.~acknowledges support from the German National Academy of Sciences Leopoldina through grant LPDS 2016-12 and from the National Science Foundation under Grant No.~DMR-2002850. We thank ISIS, STFC, UK for the Newton funding and beamtime to conduct the $\mu$SR experiments. 
\end{acknowledgments}

%===============================================
\twocolumngrid

\appendix

\section{Order parameter in band basis}\label{OrderParameterInBandBasis}

\subsection{Bloch states and microscopic representation of symmetries}
La$_7$Rh$_3$ crystallizes in the hexagonal Th$_7$Fe$_3$ structure, with space group $P6_3 mc$ (No.~186). The space group is nonsymmorphic which means that it is not possible to choose an origin such that all symmetry operations, $\{g|\vec{\tau}\} f(\vec{r}) = f(\mathcal{R}_v(g) \vec{r} + \vec{\tau})$, with $\mathcal{R}_v(g)$ denoting the vector representation of the point symmetry operation $g$, only contain Bravais-lattice translations $\vec{\tau}$; put differently, it is not possible to choose an origin such that the point group $\mathscr{P} := \{\{g|\vec{0}\} | \{g|\vec{\tau}\}  \in \mathscr{G}  \}$ is a subset of the space group $\mathscr{G}$.

The space group of La$_7$Rh$_3$ contains fractal translations along the $c$ direction corresponding to interchanging of the two alternating layers in the crystal structure. More specifically, besides lattice translations, the symmorphic symmetries are $\{E|\vec{0}\}$, $\{C_3^z|\vec{0}\}$, and $\{3\sigma_v|\vec{0}\}$ which form the group $C_{3v}$ -- a subgroup of the full point group $C_{6v}$. On top of that, there is the screw axis $\{C_6^z|\frac{1}{2}\hat{\vec{z}}\}$ (and, hence, also $\{C_2^z|\frac{1}{2}\hat{\vec{z}}\}$) and the glide planes $\{3\sigma_d|\frac{1}{2}\hat{\vec{z}}\}$.

To describe superconductivity, it is very convenient to work in second quantization. We here follow the frequently used convention and define the creation operator of an electron with crystal momentum $\vec{k}$, in atom $s$, with on-site index $\alpha$ (might just label spin or, more generally, several relevant orbitals per atom) according to
\begin{equation}
 c^\dagger_{\vec{k}\alpha s} = \frac{1}{\sqrt{N}} \sum_{\vec{R}} e^{i\vec{k}(\vec{R} + \vec{r}_s)} c^\dagger_{\vec{R} \alpha s}, \label{SecondQuantConv}
 \end{equation} 
where $N$ is the number of unit cells and $c^\dagger_{\vec{R} \alpha s}$ creates an electron in the state $\alpha$ of atom $s$ in the unit cell labeled by $\vec{R}$; the associated electronic state is located, in real space, at position $\vec{R}+\vec{r}_s$. Note that these operators satisfy the nontrivial boundary conditions
\begin{equation}
 c^\dagger_{\vec{k}+\vec{G}\alpha s} = e^{i\vec{G}\vec{r}_s} c^\dagger_{\vec{k}\alpha s} ,\qquad \vec{G}\in\text{RL}, \label{PropertyOfBoundaryCond}
\end{equation}
with RL denoting the set of reciprocal lattice vectors.
To demonstrate the consequences of these boundary conditions, consider the general superconducting mean-field Hamiltonian (with translation symmetry)
\begin{equation}
\mathcal{H}_{\text{MF}} = \sum_{\vec{k}} \left[ c^\dagger_{\vec{k}} h_{\vec{k}} c^\pdagger_{\vec{k}} + \frac{1}{2} \left( c^\dagger_{\vec{k}} \Delta_{\vec{k}} \left(c^\dagger_{-\vec{k}}\right)^T  + \text{H.c.}\right)\right], \label{GeneralMFHamiltonian}
\end{equation}
where the sum over $\vec{k}$ refers to the Brillouin zone of the Bravais lattice and we use the matrix notation implicitly containing the summation over $\alpha$ and $s$.
Shifting the summation over momentum in the mean-field Hamiltonian (\ref{GeneralMFHamiltonian}) by $\vec{G}$ and using the property (\ref{PropertyOfBoundaryCond}) of the fermionic operators, we obtain the (in general) nontrivial boundary conditions
\begin{equation}
h^\pdagger_{\vec{k}} = V^\pdagger_{\vec{G}} h^\pdagger_{\vec{k}+\vec{G}} V^\dagger_{\vec{G}}, \qquad \Delta^\pdagger_{\vec{k}} = V^\pdagger_{\vec{G}} \Delta^\pdagger_{\vec{k}+\vec{G}} V^\dagger_{\vec{G}}, \qquad \left(V_{\vec{G}}\right)_{\alpha s,\alpha' s'} = \delta_{s,s'}\delta_{\alpha,\alpha'} e^{i\vec{G}\vec{r}_s},\qquad \vec{G}\in\text{RL} . \label{BoundaryCondMicros}
\end{equation}
From the point of view of boundary conditions, it seems to be more convenient to just set $\vec{r}_s=0$ in \equref{SecondQuantConv}. However, as we will see next, the transformation properties under nonsymmorphic symmetries become more symmetric when $\vec{r}_s$ is chosen to be the position of atom $s$ in the unit cell. 

Let us consider a general space-group transformation $\{g|\vec{\tau}\}$ which acts on the real-space creation operators according to (we assume summation over repeated indices)
\begin{equation}
c^\dagger_{\vec{R}\alpha s} \, \longrightarrow \,  c^\dagger_{\vec{R}'\beta s'} \mathcal{R}_c(g,s)_{\beta\alpha} , \quad \text{with}\quad \vec{R}'+\vec{r}_{s'} = \mathcal{R}_v(g) (\vec{R} + \vec{r}_s) + \vec{\tau}. 
\end{equation}
From \equref{SecondQuantConv}, we find
\begin{align}
c^\dagger_{\vec{k}\alpha s} \, \longrightarrow \, &  \frac{1}{\sqrt{N}} \sum_{\vec{R}} e^{i\vec{k}(\vec{R} + \vec{r}_s)} c^\dagger_{\vec{R}' \beta s'} \mathcal{R}_c(g,s)_{\beta\alpha} \\
= \, &\frac{1}{\sqrt{N}} \sum_{\vec{R}} e^{i(\mathcal{R}_v(g)\vec{k})[\mathcal{R}_v(g)(\vec{R} + \vec{r}_s) +\vec{\tau}- \vec{r}_{s'}]} e^{-i(\mathcal{R}_v(g)\vec{k})(\vec{\tau}- \vec{r}_{s'})} c^\dagger_{\vec{R}' \beta s'}  \mathcal{R}_c(g,s)_{\beta\alpha} \\
= \, & \frac{1}{\sqrt{N}} \sum_{\vec{R}'} e^{i(\mathcal{R}_v(g)\vec{k})\vec{R}'} e^{-i(\mathcal{R}_v(g)\vec{k})(\vec{\tau}- \vec{r}_{s'})} c^\dagger_{\vec{R}' \beta s'}  \mathcal{R}_c(g,s)_{\beta\alpha} \\
= \, & e^{-i(\mathcal{R}_v(g)\vec{k})\vec{\tau}} \frac{1}{\sqrt{N}} \sum_{\vec{R}} e^{i(\mathcal{R}_v(g)\vec{k})(\vec{R}+\vec{r}_{s'})} c^\dagger_{\vec{R} \beta s'}  \mathcal{R}_c(g,s)_{\beta\alpha} \\
= \, & e^{-i(\mathcal{R}_v(g)\vec{k})\vec{\tau}}  c^\dagger_{\mathcal{R}_v(g)\vec{k}\beta s'}  \mathcal{R}_c(g,s)_{\beta\alpha}.
\end{align}
Introducing the multi-index $\mu=(\alpha,s)$, defining $\mathcal{R}_u(g)_{\mu'\mu}$ as the representation in the combined $(\alpha,s)$-space (which is uniquely determined by $g$ for a given crystal structure), and noting that $(\mathcal{R}_v(g)\vec{k})\vec{\tau}=\vec{k}\vec{\tau}$ can be assumed without loss of generality (only screw axes and glide planes remain after proper choice of the origin), we can write the final result as
\begin{equation}
\{g|\vec{\tau}\}: \quad c^\dagger_{\vec{k}\mu} \, \longrightarrow \,  e^{-i\vec{k}\vec{\tau}}  c^\dagger_{\mathcal{R}_v(g)\vec{k}\mu'} \mathcal{R}_u(g)_{\mu'\mu} . \label{GeneralResultForTrafoInK}
\end{equation}

Having derived the general transformation properties (\ref{GeneralResultForTrafoInK}) of the electronic operators, it is straightforward to obtain the invariance condition of the normal state Hamiltonian,
\begin{equation}
 \mathcal{R}_u(g) h_{\mathcal{R}_v^{-1}(g)\vec{k}} \mathcal{R}^\dagger_u(g) = h_{\vec{k}}, \qquad \forall\, \{g|\vec{\tau}\} \in \mathscr{G}, \label{ConstraintOnNormalStateHam}
\end{equation}
and the representation of the space group symmetries on the superconducting order parameter,
\begin{equation}
\{g|\vec{\tau}\}: \quad \Delta_{\vec{k}}\,\longrightarrow\, \mathcal{R}_u(g) \Delta_{\mathcal{R}^{-1}_v(g)\vec{k}}\mathcal{R}^T_u(g)  . \label{TrafoOfSCOP}
\end{equation}
We now see that the momentum-dependent phase factors in \equref{GeneralResultForTrafoInK} cancel and the (either lattice or fractional) translational part $\vec{\tau}$ does not show up explicitly.

\subsection{Boundary conditions in the band basis}
The weak-pairing limit, which we use to classify the possible superconducting instabilities of La$_{7}$Rh$_{3}$, was proposed in \refcite{MSS2} as a minimal description of pairing in noncentrosymmetric and spin-orbit coupled systems. The key idea is project the microscopic, matrix-valued, superconducting order parameter $\Delta_{\vec{k}}$ onto the band basis, i.e., consider instead the scalar order parameter
\begin{equation}
\widetilde{\Delta}_{\vec{k}a} = \braket{\phi_{\vec{k}a}| \Delta_{\vec{k}} T^\dagger |\phi_{\vec{k}a}},
\end{equation}
where $\phi_{\vec{k}a}$ denotes the eigenstate of the normal-state Hamiltonian $h_{\vec{k}}$ close to the Fermi surface and $T$ is the unitary part of the anti-unitary time-reversal operator $\Theta= T \mathcal{K}$, with complex-conjugation operator $\mathcal{K}$.
We refer to \refcite{MSS2}, where the same form of the transformation properties as in \equsref{ConstraintOnNormalStateHam}{TrafoOfSCOP} was used, for a detailed introduction to this approach. 

We here focus on one aspect that has not been discussed in \refcite{MSS2} but is of relevance to our discussion of symmetry-imposed zeros of $\widetilde{\Delta}_{\vec{k}a}$ at the zone-boundary: despite the nontrivial boundary conditions in \equref{BoundaryCondMicros}, $\widetilde{\Delta}_{\vec{k}a}$ satisfies periodic boundary conditions,
\begin{equation}
 \widetilde{\Delta}_{\vec{k}a} =  \widetilde{\Delta}_{\vec{k}+\vec{G}a}, \qquad \vec{G}\in\text{RL}, \label{BoundaryCondOrderParam}
\end{equation}
as we prove next. We first note that \equref{BoundaryCondMicros} implies for non-degenerate bands
\begin{equation}
\phi_{\vec{k}+\vec{G}a} = e^{i\varphi_{\vec{k}a}} V^\dagger_{\vec{G}}\phi_{\vec{k}a}, \label{ConditionOnWFs}
\end{equation}
where $e^{i\varphi_{\vec{k}a}}$ is an arbitrary phase factor and the association of both wavefunctions with the same band index is purely a convention which, however, is very natural: if the Fermi surface $a$ does not cross the boundary of the Brillouin zone, \equref{ConditionOnWFs} means that we just give the equivalent band outside the Brillouin zone the same label; if it crosses the zone boundary, this convention will be required in order to give the same label $a$ to all momentum points on a connected Fermi surface.

Applying \equref{ConditionOnWFs} and the boundary condition of the superconducting order parameter in \equref{BoundaryCondMicros}, we find
\begin{equation}
\widetilde{\Delta}_{\vec{k}+\vec{G} a} = \braket{\phi_{\vec{k}a}| V^\pdagger_{\vec{G}} V^\dagger_{\vec{G}} \Delta_{\vec{k}} V^\pdagger_{\vec{G}} T^\dagger  V^\dagger_{\vec{G}} |\phi_{\vec{k}a}} = \widetilde{\Delta}_{\vec{k}a},
\end{equation}
where we have used that $V^\pdagger_{\vec{G}}$ and $T^\dagger$ commute which is a consequence of time-reversal being local ($T_{\alpha s,\alpha' s'} = \widetilde{T}_{\alpha,\alpha'}(s) \delta_{s,s'}$).

\begin{table}[t]
\begin{center}
\caption{Possible pairing states in La$_{7}$Rh$_{3}$ classified according to the IRs of the point group $C_{6v}$. Here, $X$, $Y$, and $Z$ are continuous, real-valued functions on the Brillouin zone which are odd under $\vec{k} \rightarrow -\vec{k}$ and transform under $C_{6v}$ as $k_x$, $k_y$, and $k_z$, respectively, with the six-fold rotation symmetry along $k_z$ and the $k_x$-, $k_y$-axes oriented normal to two of the six mirror planes. We have already taken into account the constraint, $\varphi_{na}^\mu({\vec{k}})=\varphi_{na}^\mu({-\vec{k}})$, of the scalar basis functions resulting from TRS and Fermi statistics \cite{MSS2}. This is why we have omitted, e.g., $(X,Y)$ as basis functions of $E_1$.}
\label{PossiblePairingStates}
\begin{tabular} {ccc} \hline \hline
\hspace{0.6em}    IR  \hspace{0.6em} & \hspace{0.6em} $d_n$ \hspace{0.6em} & \hspace{0.6em} Symmetry/leading basis functions $\varphi_{na}^\mu({\vec{k}})$ \hspace{0.6em}\\ \hline
 $A_1$      & 1                 & $1$, $X^2+Y^2$, $Z^2$  \\
$A_2$         & 1    & $XY(3X^2-Y^2)(3Y^2-X^2)$ \\
$B_1$        & 1      & $XZ(3Y^2-X^2)$\\
$B_2$          & 1    & $YZ(3X^2-Y^2)$ \\ \hline
$E_1$          & 2     & $(XZ,YZ)$\\
$E_2$          & 2    & $(X^2-Y^2,2XY)$\\ \hline \hline
 \end{tabular}
\end{center}
\end{table}

\subsection{Basis functions}
In \tableref{PossiblePairingStates}, we summarize the scalar basis functions $\varphi^\mu_{na}(\vec{k})$ for the different IRs of the point group $C_{6v}$ of La$_{7}$Rh$_{3}$ that determine the weak-pairing order parameter according to $\widetilde{\Delta}_{\vec{k}a} = \sum_{\mu=1}^{d_n} \eta_\mu\varphi^\mu_{na}(\vec{k})$. As mentioned in the main text, a Ginzburg-Landau analysis beyond quadratic order shows that only the three symmetry-inequilvalent combinations $(\eta_1,\eta_2)=(1,0),\,(0,1),\,(1,i)$ are possible in the case of the two 2D IRs $E_1$ and $E_2$.

\section{Symmetry protected zeros}\label{SymmetryProtectedZeros}

The symmetry-protected, i.e., non-accidental, nodes of the two pairing states that break TRS are shown in \figref{SymmProtectZeros}.
To understand how these zeros emerge, we first note that, for both of the two TRS-breaking states transforming under $n=E_1$ and $n=E_2$, the superconducting order parameter has the form $\Delta_{\vec{k} a} = \varphi_{n a}^{(1)}(\vec{k})+i \varphi_{n a}^{(2)}(\vec{k})$ leading to the gap function \cite{MSS2}
\begin{equation}
|\widetilde{\Delta}_{\vec{k} a}| = \sqrt{\left(\varphi_{n a}^{(1)}(\vec{k})\right)^2+\left(\varphi_{n a}^{(2)}(\vec{k})\right)^2}, \qquad n=E_1,E_2.
\end{equation}
Consequently, the gap vanishes at $\vec{k} = \vec{k}_0$ if and only if $\varphi_{n a}^{(1)}(\vec{k}_0)=\varphi_{n a}^{(2)}(\vec{k}_0)=0$. In the following, we analyze where this condition is met, discussing the $E_1(1,i)$ and $E_2(1,i)$ states separately.

\subsection{Order parameter $E_1(1,i)$}
For this IR, the basis functions $\varphi_{E_1 a}^{(1)}(\vec{k})$ and $\varphi_{E_1 a}^{(2)}(\vec{k})$ have to transform as $k_x$ and $k_y$ under $C_{6v}$. For any $\vec{k}=\vec{k}^*$ with $\mathcal{R}_v(C^z_n)\vec{k}^*=\vec{k}^*$ (fixed point), $n=2,3,6$, it thus holds
\begin{equation}
\begin{pmatrix} \varphi_{E_1 a}^{(1)}(\vec{k}^*) \\ \varphi_{E_1 a}^{(2)}(\vec{k}^*) \end{pmatrix} = \begin{pmatrix} \varphi_{E_1 a}^{(1)}(\mathcal{R}^{-1}_v(C^z_n)\vec{k}^*) \\ \varphi_{E_1 a}^{(2)}(\mathcal{R}^{-1}_v(C^z_n)\vec{k}^*) \end{pmatrix}  = \mathcal{R}^{xy}_v(C^z_n)\begin{pmatrix} \varphi_{E_1 a}^{(1)}(\vec{k}^*) \\ \varphi_{E_1 a}^{(2)}(\vec{k}^*) \end{pmatrix}, \label{RotationalBeh}
 \end{equation} 
where $\mathcal{R}^{xy}_v(C^z_n)$ is the vector representation of $C^z_n$ in the xy-plane. \equref{RotationalBeh} implies $\varphi_{E_1 a}^{(1)}(\vec{k}^*)=\varphi_{E_1 a}^{(2)}(\vec{k}^*)=0$ for any fixed point $\vec{k}^*$ of $C^z_2$, $C^z_3$, or $C^z_6$. This leads to the red lines in \figref{SymmProtectZeros}(a) through the $\Gamma$, K, and M points.

The second generator of the point group, one of the reflections, say $\sigma_{xz}:\,(k_x,k_y,k_z)\rightarrow (k_x,-k_y,k_z)$, does not yield symmetry-protected nodes: While it implies $\varphi_{E_1 a}^{(2)}(k_x,0,k_z)=0$, the other component can be non-zero on the $k_x$-$k_z$ plane.

As already discussed in the main text, the constraint $\varphi_{E_1a}^\mu({\vec{k}})=\varphi_{E_1a}^\mu({-\vec{k}})$ combined with $C^z_2$ leads to
\begin{equation}
\varphi_{E_1 a}^{\mu}(k_x,k_y,k_z) = -\varphi_{E_1 a}^{\mu}(-k_x,-k_y,k_z) = -\varphi_{E_1 a}^{\mu}(k_x,k_y,-k_z). \label{ConstrResultFromWP}
\end{equation}
Together with the boundary conditions (\ref{BoundaryCondOrderParam}), we conclude
\begin{equation}
\varphi_{E_1 a}^{\mu}(k_x,k_y,k_z) = 0, \qquad k_z=0,\pi/c, \label{VanishingBasisFunc}
\end{equation}
leading to the two distinct nodal planes in \figref{SymmProtectZeros}(a). We emphasize that generic points in the planes defined by $k_z=0$ and $k_z=\pi/c$ are no high-symmetry points, i.e., are only invariant under the identity transformation of the point group (the little point group of these momenta is trivial, only admitting 1D IRs). Consequently, our assumption of non-degenerate bands is (in general) expected to hold for almost all momenta in these planes and \equref{VanishingBasisFunc} leads to line nodes for Fermi surfaces crossing $k_z=0,\pi/c$. Note that the presence of these line nodes is related to the selection rule derived in \refcite{MSS2} which is based on the observation that $\varphi_{E_1a}^\mu({\vec{k}})=\varphi_{E_1a}^\mu({-\vec{k}})$ does not allow for a superconducting gap in a 2D system if the order parameter is odd under a two-fold rotation perpendicular to the plane of the system.

%\begin{figure}
%\includegraphics[width=0.6\columnwidth]{ProtectedZeros}
%\caption{\label{SymmProtectZeros}The location of symmetry-imposed zeros of the weak-pairing gap $|\widetilde{\Delta}_{\vec{k} a}|$ in the Brillouin zone are indicated in red for the TRS-breaking candidate states $E_1(1,i)$ and $E_2(1,i)$ in (a) and (b), respectively. Note that we here assume that the bands are non-degenerate in the entire Brillouin zone which is necessary for the weak-pairing description to apply. This is a good assumption for generic momentum points (invariant only under the identity operation of the point group), including generic points in the planes $k_z=0,\pi/c$. However, it will not necessarily be satisfied at the high-symmetry lines parallel to the $k_z$ axis and going through the indicated high-symmetry points.}
%\end{figure}

\subsection{Order parameter $E_2(1,i)$}
Let us now perform the same analysis for $E_2$. Here $\varphi_{E_2 a}^{(1)}(\vec{k})$ and $\varphi_{E_2 a}^{(2)}(\vec{k})$ transform as $k^2_x-k_y^2$ and $2k_x k_y$ under $C_{6v}$ which leads to important differences. First of all, the transformation behavior under $C^z_6$ implies
\begin{equation}
 \begin{pmatrix} \varphi_{E_2 a}^{(1)}(\mathcal{R}^{-1}_v(C^z_6)\vec{k}) \\ \varphi_{E_2 a}^{(2)}(\mathcal{R}^{-1}_v(C^z_6)\vec{k}) \end{pmatrix}  = \mathcal{R}^{xy}_v(C^z_3)\begin{pmatrix} \varphi_{E_2 a}^{(1)}(\vec{k}) \\ \varphi_{E_2 a}^{(2)}(\vec{k}) \end{pmatrix}.
\end{equation}
Consequently, the two components only have to vanish at fixed points, $\vec{k}=\vec{k}^*$ with $C^z_n\vec{k}^* = \vec{k}^*$, of $n=6$-fold and $n=3$-fold rotation [but no constraint for fixed points of $C^z_2=(C^z_6)^3$ as $(\mathcal{R}_v(C^z_3))^3=\mathbbm{1}$]. Consequently, there are only line nodes through the $\Gamma$ and K points (but not through the M point), see \figref{SymmProtectZeros}(b).

Exactly as for $E_1$, the additional reflection symmetries do not impose further zeros. Furthermore, being even under $C^z_2$, the analog of \equref{ConstrResultFromWP} just reads as $\varphi_{E_2 a}^{\mu}(k_x,k_y,k_z) = \varphi_{E_2 a}^{\mu}(k_x,k_y,-k_z)$ and, hence, does not impose any nodal planes.

\section{Signatures of spin fluctuations and Wilson ratio}\label{WilsonRatio}
In this part of the appendix, we discuss experimental evidence for the presence of sizable electronic correlations, in particular, spin fluctuations, which underpin the relevance of electron-electron interactions for the superconducting instability.

\begin{figure}[b]
\begin{center}
\includegraphics[width=9.cm]{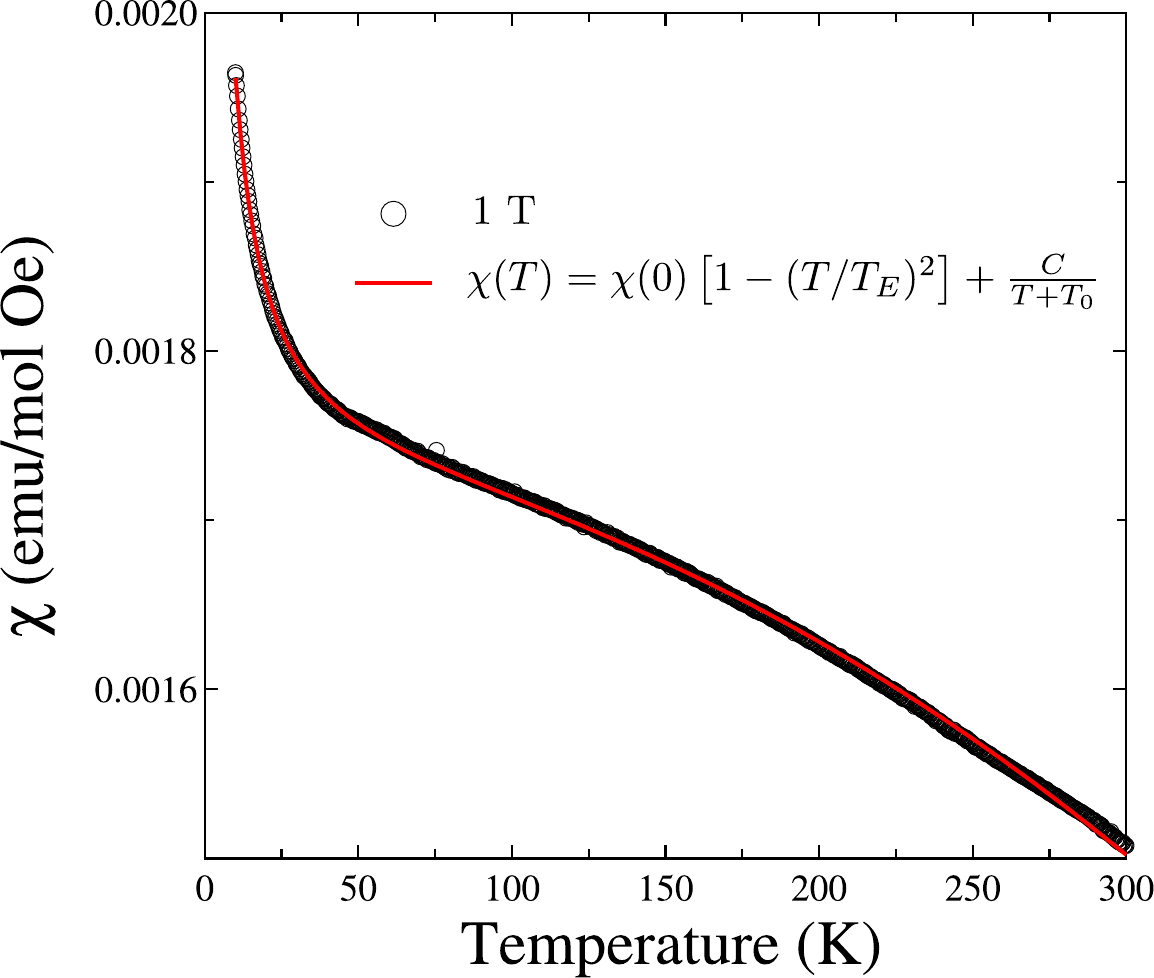}
\caption{\label{Fig1:cp} Temperature dependence of the magnetic susceptibility measured at 1 T.  The solid line is a fit to the indicated function form \cite{SuceptFit} for $\chi(T)$.}
\end{center}
\end{figure}

%From previous work:
Before presenting new data, we briefly discuss indications from previous literature.
In \refcite{PP}, signs of spin fluctuations in La$_7$Rh$_3$ have been reported based on the observation that the susceptibility shows a broad maximum at around 150 K -- very similar to that of pure Pd metal, which is a typical example of an ``incipient ferromagnet''. We point out that similar behavior was also observed, e.g., in TiBe$_2$ \cite{TiBe2}, (Ca,Sr)$_{2}$RuO$_{4}$ \cite{CaSrRuO}, and MnSi \cite{MnSi} which are known examples of systems with spin fluctuations. 

As argued in \refcite{PP}, we can also get evidence for spin fluctuations by looking at the temperature dependence of resistivity, $\rho(T)$, near $T_{c}$.
It has been reported that the La$_{7}$X$_{3}$ (X = Ir, Ru) \cite{DFT,PP} series of compounds, isomorphic to La$_{7}$Rh$_{3}$, exhibits low-temperature $\rho(T)$, which is similar to YCo$_{2}$, an archetype of a spin-fluctuating compound. 

In order to quantify the importance of electronic correlations, we have determined the Wilson ratio from our specific heat data [see \figref{Fig1:xrd}(b) of the main text] and our magnetic susceptibility measurements, see \figref{Fig1:cp}.
The Wilson ratio is defined as the dimensionless quantity \cite{WilsonRatio}
\begin{equation}
R_{w} = \frac{4 \pi^{2}{k_{B}^{2}} \chi(0)}{3 (g \mu_{B})^{2} \gamma_{n}},
\end{equation}
where $\chi(0)$ is the Pauli susceptibility of the electrons at zero temperature, $\gamma_{n}$ is the specific heat coefficient, $g$ and $\mu_{B}$ are the Land\'e factor and the Bohr magneton, respectively. 
The Wilson ratio $R_{w}$ reveals how strong electronic correlations are; it holds $R_{w}$ $\approx$ 1 for the non-interacting electron gas, $R_{w}$ $\approx$ 1-2 for weakly interacting Fermi liquids, and larger than 2 for strongly correlated systems.

We have fitted the indicated phenomenological functional form for $\chi(T)$ to the measured susceptibility to extract $\chi(0)$ =  0.001713 emu/mol Oe. Using $\gamma_{n}$= 42.06 mJ/mol K$^{2}$ extracted from the heat capacity data, we estimated the value of the Wilson ratio to be $R_w  = 2.966$, indicating strong electronic correlations in La$_{7}$Rh$_{3}$.

% Summary: 
Together with the time-reversal-symmetry-breaking at the superconducting transition, these observations, in conclusion, suggest that superconductivity in La$_{7}$Rh$_{3}$ is crucially driven by electron-electron interactions.

\end{document}